\documentclass[aps,prd,10pt,twocolumn,nofootinbib]{revtex4-2}

\usepackage{graphicx}
\usepackage{amsmath}
\usepackage{amssymb}
\usepackage{bm}
\usepackage{bbm}
\usepackage{color}
\usepackage{slashed}
\usepackage[colorlinks=true,linkcolor=blue,citecolor=blue, urlcolor=blue]{hyperref}

\allowdisplaybreaks

\begin{document}

\title{Next-to-leading-order QCD study of $e^+e^- \to \eta_c + c\bar{c}$ at $B$ factories}

\author{Lu-Yao Zhang}
\email{luyaozhang@stu.cqu.edu.cn}
\author{Xu-Chang Zheng}
\email{zhengxc@cqu.edu.cn}
\author{Xing-Gang Wu}
\email{wuxg@cqu.edu.cn}

\affiliation{Department of Physics, Chongqing Key Laboratory for Strongly Coupled Physics, Chongqing University, Chongqing 401331, People's Republic of China}

\begin{abstract}

We study the process $e^+e^- \to \gamma^\ast \to \eta_c + c\bar{c} + X$ at next-to-leading order (NLO) in QCD within the framework of nonrelativistic QCD (NRQCD) factorization, retaining the color-singlet contribution at leading order in the heavy-quark relative velocity $v$. At the $e^+e^-$ collision energy $\sqrt{s} = 10.6~\mathrm{GeV}$, with $m_c = 1.5~\mathrm{GeV}$ and $\mu_R = 2m_c$, the direct cross section increases from $58.7~\mathrm{fb}$ at leading order (LO) to $118~\mathrm{fb}$ at NLO, corresponding to a $K$ factor of approximately $2.0$. Including the radiative feed-down contributions of the $J/\psi$ and $h_c(1P)$ states yields an estimated prompt production rate of $131~\mathrm{fb}$. The NLO corrections substantially enhance both the angular and momentum distributions, while the momentum spectrum retains a similar overall profile. These results establish an NLO color-singlet baseline for $\eta_c$ production associated with open charm at Belle II, complementary to studies of charmonium production in hadronic collisions.

\end{abstract}

%\pacs{12.38.Bx, 13.87.Fh, 13.66.Bc}

\maketitle

\section{Introduction}
\label{secIntro}

Heavy quarkonium production probes both perturbative QCD and the nonperturbative formation of heavy-quark bound states~\cite{Brambilla:2005, Brambilla:2010cs, Andronic:2015wma, Lansberg:2019adr}. The color-singlet model~\cite{Chang:1979nn, Berger:1980ni, Baier:1981uk} describes production through a heavy-quark pair with the quantum numbers of the observed quarkonium. Nonrelativistic QCD (NRQCD) factorization~\cite{Bodwin:1994jh} embeds this contribution, together with color-octet channels, in a systematic expansion in the strong coupling $\alpha_s$ and the relative heavy-quark velocity $v$. It separates perturbative short-distance coefficients (SDCs) from long-distance matrix elements (LDMEs), whose universality requires the same matrix elements to describe different production processes for a given quarkonium state.

A central challenge is to describe production rates and polarization with a consistent set of LDMEs. Measurements of prompt $J/\psi$ and $\psi(2S)$ polarization at the LHC disfavor strong transverse polarization, including at large transverse momentum~\cite{CMS:2013pol, CMS:2024pol}. Next-to-leading-order (NLO) analyses using different data selections have obtained substantially different color-octet LDMEs and polarization predictions~\cite{Butenschoen:2012pol, Chao:2012pol, Gong:2013pol}. The interplay between yields and polarization therefore makes comparisons across observables and production environments essential for assessing LDME universality.

The pseudoscalar $\eta_c$ provides an important constraint on this problem. Heavy-quark spin symmetry relates its LDMEs to those of its spin-triplet partner $J/\psi$; in particular, the $\eta_c$ ${}^3S_1^{[8]}$ LDME is related to the $J/\psi$ ${}^1S_0^{[8]}$ LDME up to relative corrections of order $v^2$~\cite{Han:2015etac, Zhang:2015etac}. Measurements of prompt $\eta_c$ hadroproduction by LHCb~\cite{LHCb:2015etac, LHCb:2020etac, LHCb:2024etac} thus also constrain the interpretation of $J/\psi$ production. NLO studies found that the color-singlet contribution alone could approximately saturate the initial LHCb data in the measured kinematic region, whereas adding color-octet contributions obtained from several $J/\psi$ LDME extractions substantially overestimated the cross section~\cite{Butenschoen:2015etac, Han:2015etac}. Other analyses incorporating the $\eta_c$ constraints obtained compatible descriptions of the $\eta_c$ and $J/\psi$ yields and selected polarization data~\cite{Han:2015etac, Zhang:2015etac}.

The extended LHCb measurement at 13 TeV further illustrates this tension: the color-singlet prediction describes the data reasonably well for $p_T>8~\mathrm{GeV}$ but overestimates them at lower $p_T$, while the specific color-singlet-plus-color-octet prediction compared with the data is too large throughout the measured range~\cite{LHCb:2024etac}. These findings constrain the allowed LDMEs and motivate independent tests of the color-singlet contribution in other production environments.

Electron--positron collisions offer such an opportunity, with well-defined initial states and no dependence on hadronic parton distributions. Measurements at the $B$ factories, operating near $\sqrt{s}=10.6~\mathrm{GeV}$, revealed substantial differences between leading-order (LO) predictions and the observed rates for double-charmonium and associated open-charm production~\cite{Belle:2002ric, Belle:2004, BaBar:2005nic, Braaten:2002fzi, Hagiwara:2003bc, Liu:2002wq}. Higher-order effects play different roles in these channels. For $e^+e^-\to J/\psi+\eta_c$, both QCD radiative corrections~\cite{Zhang:2006, Gong:2009} and relativistic corrections~\cite{He:2007te} enhance the prediction. For $e^+e^-\to J/\psi+c\bar{c}$, NLO QCD calculations substantially increase the color-singlet cross section and, together with feed-down and consistent input parameters, improve the description of the measured rate~\cite{Zhang:2006ay, Gong:2009ng, Huang:2024zni}.

The process $e^+e^-\to\gamma^\ast\to\eta_c+c\bar{c}+X$, where $X$ denotes additional QCD radiation, offers a complementary probe of spin-singlet production with an associated heavy-quark pair. In the standard NRQCD velocity counting, the ${}^1S_0^{[1]}$ channel gives the leading contribution, while the leading color-octet terms enter at relative order $v^4$~\cite{Bodwin:1994jh, Butenschoen:2015etac}. Their numerical weights also depend on the process-dependent short-distance coefficients. Our aim is to determine the normalization and kinematic dependence of the leading color-singlet term at NLO, providing a quantitative reference for subsequent assessments of higher-order terms in the velocity expansion. This contribution involves a single LDME, fixed through the quarkonium wavefunction independently of color-octet production fits.

The LO color-singlet cross section for this open-charm channel is known and starts at order $\alpha^2\alpha_s^2$~\cite{Liu:2004}. By comparison, in $\eta_c$ production with light hadrons, color-octet channels can benefit from a lower starting order in $\alpha_s$ than the color-singlet channel~\cite{Gong:2016lh}. These processes therefore provide complementary information on quarkonium formation. Related NLO studies of pseudoscalar quarkonium associated with a heavy-quark pair concern photon fusion~\cite{Yang:2022gg} and $Z$-boson decay~\cite{Zheng:2022eta}, which involve different initial-state couplings. To the best of our knowledge, the NLO QCD correction to the single-virtual-photon open-charm channel considered here has not previously been reported.

In this paper we calculate the color-singlet ${}^1S_0^{[1]}$ contribution through order $\alpha^2\alpha_s^3$, at leading order in  $v$. The prediction includes both virtual corrections and real-gluon emission and is inclusive with respect to the additional radiation. We study the dependence of the total cross section on the charm-quark mass $m_c$ and the renormalization scale $\mu_R$, and present the $\eta_c$ angular and momentum distributions at the central inputs.

For comparison with future measurements, we also estimate the prompt rate and distributions by including the radiative feed-down contributions from $J/\psi$ and $h_c(1P)$. The parent-production calculations and treatment of the decay kinematics are specified in Sec.~\ref{secNumer}. The total cross sections cover the full phase space, without detector acceptance cuts.

The rest of this paper is organized as follows. Section~\ref{secLO} specifies the LO amplitudes and normalization conventions. Section~\ref{secNLO} describes the virtual and real NLO corrections. Numerical results for direct and prompt production are presented in Sec.~\ref{secNumer}, and Sec.~\ref{secCon} gives a brief summary.

\section{Leading-Order Calculation}
\label{secLO}

In this section, we calculate the production of $\eta_c$ in association with an open charm-anticharm pair through a virtual photon. In NRQCD factorization, the physical cross section is expressed in terms of perturbative SDCs and nonperturbative LDMEs.

For the LO process, the NRQCD factorization formula reads
\begin{align}
&d\sigma(e^+ e^- \to \eta_c + c\bar{c}) \notag\\
&=\sum_n d\hat{\sigma}(e^+ e^- \to (c\bar{c})[n]+c\bar{c})\langle{\cal O}^{\eta_c}(n)\rangle.
\end{align}
Here, $n={}^{2S+1}L_J^{[1,8]}$ labels the spin, orbital angular momentum, and color of the intermediate $c\bar{c}$ pair. We retain the color-singlet state $n={}^1S_0^{[1]}$ at leading order in the relative velocity $v$. The quantity $d\hat\sigma$ is the SDC obtained by perturbative matching, while $\langle \mathcal{O}^{\eta_c}(n)\rangle$ describes the formation of the physical quarkonium. We use the projector and NRQCD matrix-element normalization of Ref.~\cite{Artoisenet:2014lpa}, as specified below. The LDME can be determined from experimental data, lattice QCD~\cite{lattice1, lattice2}, or potential models~\cite{potenmodel1, potenmodel2, potenmodel3, potenmodel4}; here it is expressed in terms of the radial wavefunction at the origin.

\begin{figure}[htb]
\centering
\includegraphics[width=0.45\textwidth]{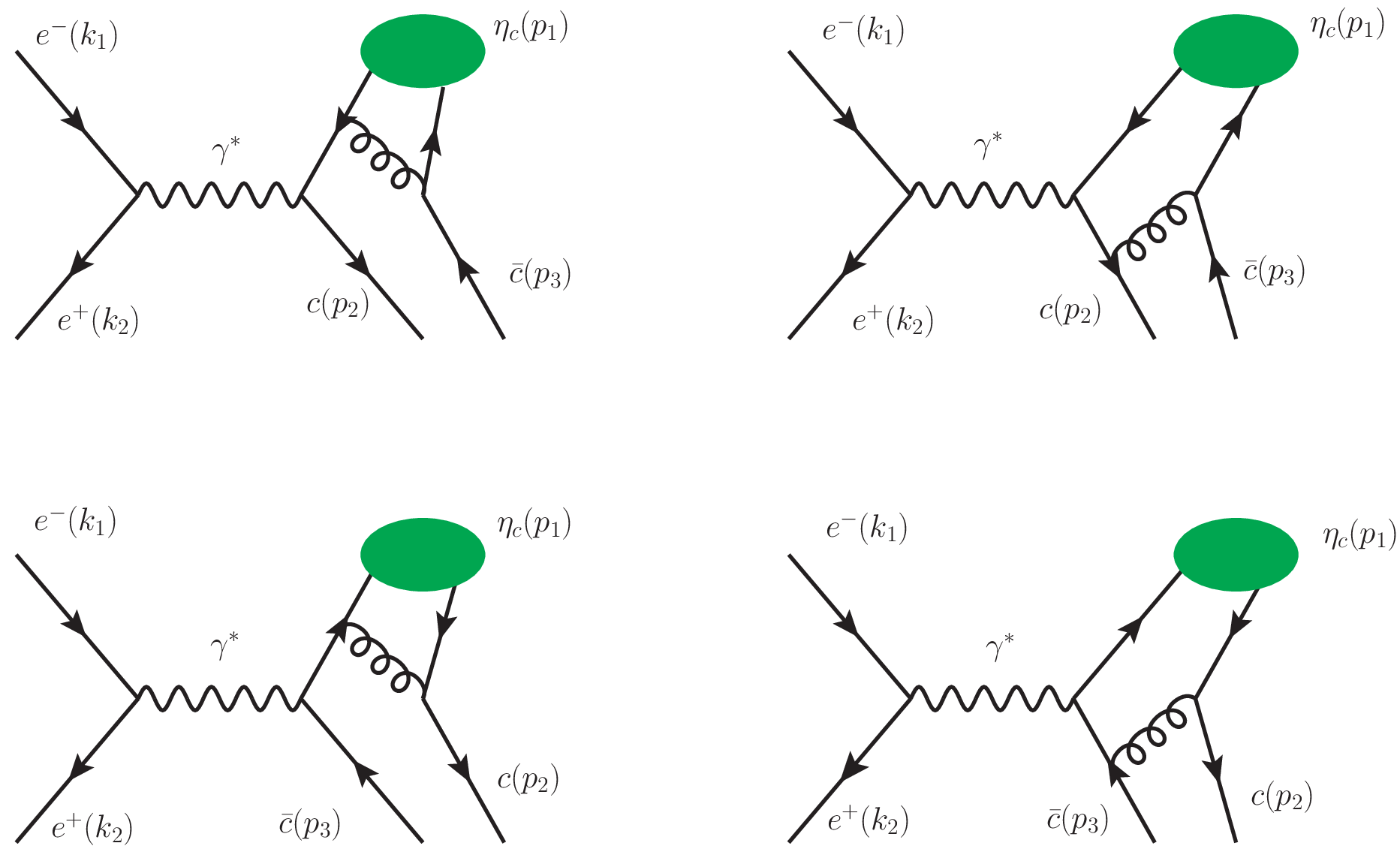}
\caption{LO diagrams for $e^-(k_1)+e^+(k_2)\to(c\bar c)[{}^1S_0^{[1]}](p_1)+c(p_2)+\bar c(p_3)$.
} \label{tree}
\end{figure}

The LO diagrams for producing the projected free pair $(c\bar c)[n]$ are shown in Fig.~\ref{tree}. At leading order in $v$, its constituents each carry momentum $p_1/2$, with $p_1^2=4m_c^2$ and $p_2^2=p_3^2=m_c^2$. The amplitudes below describe this free pair and do not include the quarkonium wavefunction. We use the electromagnetic coupling $e$, with $e^2=4\pi\alpha$, and the strong coupling $g_s$, with $g_s^2=4\pi\alpha_s$. The fundamental color generators are denoted by $T^a$. The LO amplitude is $\mathcal M_{\rm LO}=\mathcal M_1+\mathcal M_2+\mathcal M_3+\mathcal M_4$, where
\begin{align}
	i\mathcal{M}_1 ={}& \frac{-i}{(p_1/2 + p_2)^2 + i\epsilon} \frac{-i}{(p_1 + p_2 + p_3)^2 + i\epsilon} \notag\\*
	& \times \bar{v}(k_2) (i e \gamma^\mu) u(k_1) \bar{u}(p_2) (-i g_s \gamma^\nu T^a) \notag\\*
	& \times \Pi_0 \Lambda_1 (-i g_s \gamma_\nu T^a) \frac{i}{\slashed{p}_1 + \slashed{p}_2 - m_c + i\epsilon} \notag\\*
	& \times \left( -\frac{2}{3} i e \gamma_\mu \right) v(p_3),\\
	i\mathcal{M}_2 ={}& \frac{-i}{(p_1/2 + p_2)^2 + i\epsilon} \frac{-i}{(p_1 + p_2 + p_3)^2 + i\epsilon} \notag\\*
	& \times \bar{v}(k_2) (i e \gamma^\mu) u(k_1) \bar{u}(p_2) (-i g_s \gamma^\nu T^a) \notag\\*
	& \times \Pi_0 \Lambda_1 \left( -\frac{2}{3} i e \gamma_\mu \right) \frac{i}{-\slashed{p}_1/2 - \slashed{p}_2 - \slashed{p}_3 - m_c + i\epsilon} \notag\\*
	& \times (-i g_s \gamma_\nu T^a) v(p_3),\\
	i\mathcal{M}_3 ={}& \frac{-i}{(p_1/2 + p_3)^2 + i\epsilon} \frac{-i}{(p_1 + p_2 + p_3)^2 + i\epsilon} \notag\\*
	& \times \bar{v}(k_2) (i e \gamma^\mu) u(k_1) \bar{u}(p_2) \left( -\frac{2}{3} i e \gamma_\mu \right) \notag\\*
	& \times \frac{i}{-\slashed{p}_1 - \slashed{p}_3 - m_c + i\epsilon} (-i g_s \gamma^\nu T^a) \Pi_0 \Lambda_1 \notag\\*
	& \times (-i g_s \gamma_\nu T^a) v(p_3),
\end{align}
\begin{align}
	i\mathcal{M}_4 = & \frac{-i}{(p_1/2 + p_3)^2 + i\epsilon} \frac{-i}{(p_1 + p_2 + p_3)^2 + i\epsilon} \notag\\
	& \times \bar{v}(k_2) (i e \gamma^\mu) u(k_1) \bar{u}(p_2) (-i g_s \gamma^\nu T^a) \notag\\
	& \times \frac{i}{\slashed{p}_1/2 + \slashed{p}_2 + \slashed{p}_3 - m_c + i\epsilon} \left( -\frac{2}{3} i e \gamma_\mu \right) \Pi_0 \Lambda_1 \notag\\
	& \times (-i g_s \gamma_\nu T^a) v(p_3),
\end{align}
	where $\Pi_0$ denotes the spin projection operator for the spin-singlet $S$-wave state,
	\begin{equation}
		\Pi_0 = \frac{1}{(2m_c)^{3/2}} (\slashed{p}_1/2 - m_c) \gamma^5 (\slashed{p}_1/2 + m_c),
	\end{equation}
	and $\Lambda_1$ represents the color-singlet projector,
	\begin{equation}
		\Lambda_1 = \frac{\mathbf{1}}{\sqrt{N_c}},
	\end{equation}
with $\mathbf{1}$ the color identity and $N_c=3$. As in Ref.~\cite{Artoisenet:2014lpa}, the projected pair is relativistically normalized as a particle of mass $2m_c$. The corresponding perturbative and physical NRQCD matrix elements are
\begin{align*}
\langle {\cal O}^{c\bar c[n]}(n)\rangle &= 2N_c,\\
\langle {\cal O}^{\eta_c}(n)\rangle &= \frac{N_c}{2\pi}|R_{\eta_c}(0)|^2,
\end{align*}
where the second relation uses vacuum saturation at the retained order in $v$. Writing the rate computed from the projected amplitudes as $d\sigma_{\rm pair}$, matching gives
\begin{align*}
d\hat\sigma &= \frac{d\sigma_{\rm pair}}{2N_c},\\
d\sigma &= d\hat\sigma\,\langle {\cal O}^{\eta_c}(n)\rangle
          = \frac{|R_{\eta_c}(0)|^2}{4\pi}\,d\sigma_{\rm pair}.
\end{align*}
In dimensional regularization at zero relative momentum, this perturbative LDME receives no NLO correction~\cite{Artoisenet:2014lpa}, so the same matching applies at NLO.

The unpolarized LO SDC is therefore
\begin{equation}
d\hat{\sigma}_{\text{LO}} = \frac{1}{2N_c}\frac{1}{4}\frac{1}{2s} \sum |\mathcal{M}_{\text{LO}}|^2 d\Phi_3,
\end{equation}
where $s=(k_1+k_2)^2$, the electron mass is neglected, and $1/(2s)$ is the flux factor. The sum includes the initial lepton spins and the spins and colors of the open charm pair, while $1/4$ averages over the initial spins. The spin and color of $(c\bar c)[n]$ have already been fixed by the projectors. The three-body phase space treats the projected pair as one particle of momentum $p_1$ and reads
\begin{equation}
	d\Phi_3 = (2\pi)^4 \delta^4 \left(k_1 + k_2 - \sum_{f=1}^3 p_f \right) \prod_{f=1}^3 \frac{d^3\mathbf{p}_f}{(2\pi)^3 2p_f^0},
\end{equation}
where $p_f$ denotes the four-momenta of the final-state particles.

\section{Next-to-Leading-Order Calculation}
\label{secNLO}

At order $\alpha^2\alpha_s^3$, the direct cross section receives contributions from one-loop corrections to the Born process and from real-gluon emission. Both contributions are needed for an infrared-finite inclusive prediction.

At NLO, we calculate the virtual and real corrections to production of the same projected free pair and divide by its perturbative LDME $2N_c$. The resulting SDC is
\begin{equation}
\hat{\sigma}_{\rm NLO} = \hat{\sigma}_{\rm LO} + \hat{\sigma}_{\rm Virtual} + \hat{\sigma}_{\rm Real},
\end{equation}
where $\hat{\sigma}_{\rm Virtual}$ and $\hat{\sigma}_{\rm Real}$ are the corresponding corrections to the SDC. Multiplication by $\langle {\cal O}^{\eta_c}(n)\rangle$ gives the physical direct-production cross section. The amplitudes in both contributions use the free-pair projectors defined in Sec.~\ref{secLO}.

\subsection{Virtual corrections}

\begin{figure}[htb]
    \centering
	\includegraphics[width=0.45\textwidth]{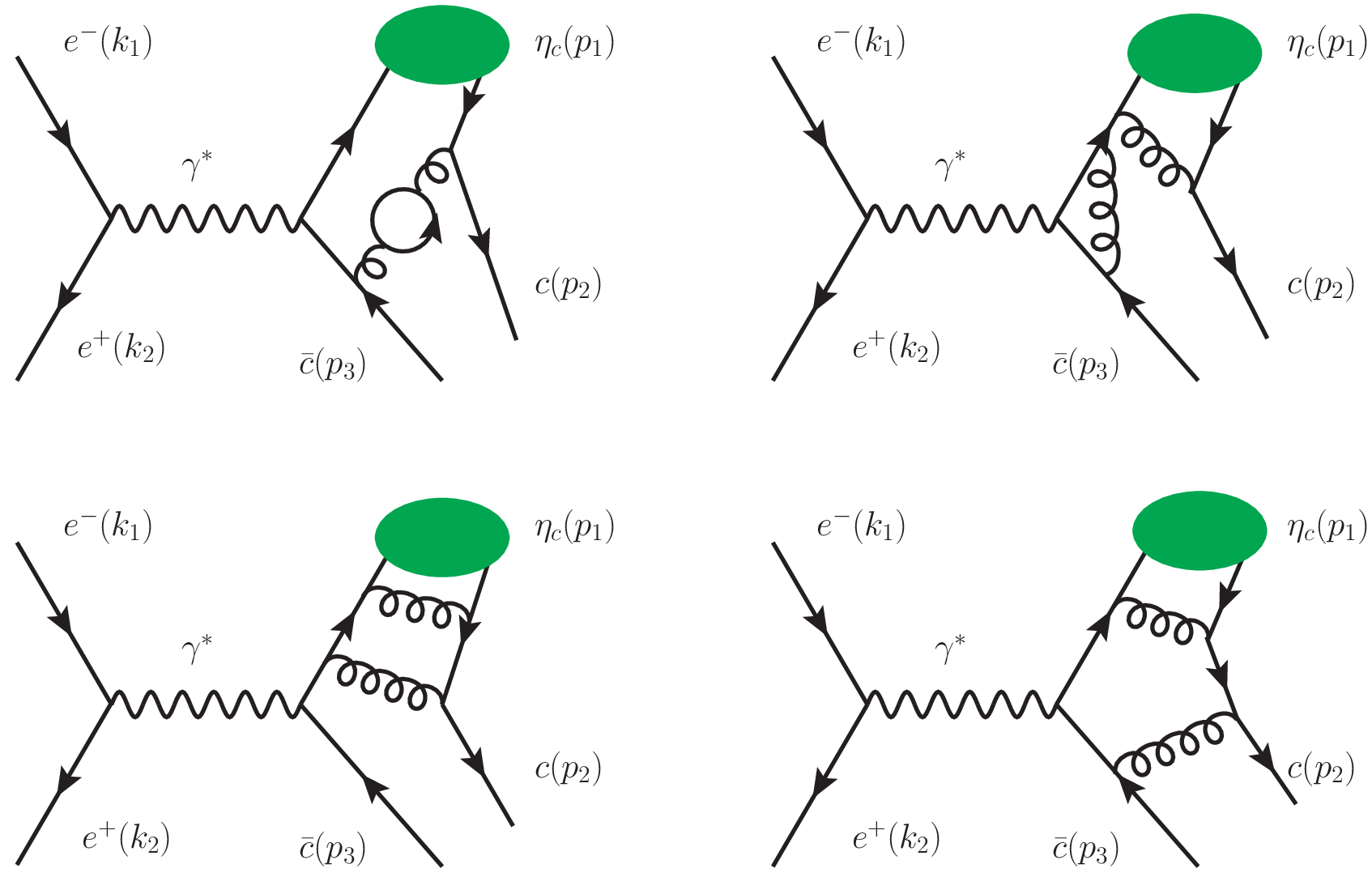}
	\caption{Representative one-loop diagrams for $e^+e^-\to(c\bar c)[{}^1S_0^{[1]}]+c\bar c$.
	} \label{vir}
\end{figure}

The virtual contribution is obtained from the interference of the LO amplitude with the one-loop amplitude, including counterterms. The loop diagrams comprise 28 self-energy, 20 vertex, 18 box, and 6 pentagon diagrams; representative diagrams are shown in Fig.~\ref{vir}. The differential contribution to the SDC is
\begin{align}
d\hat{\sigma}_{\rm Virtual} ={}& \frac{1}{2N_c}\frac{1}{4}\frac{1}{2s}\notag\\
&\times\sum 2\text{Re}(\mathcal{M}_{\text{LO}}^*\mathcal{M}_{\text{Virtual}})d\Phi_3,
\end{align}
where $\mathcal M_{\rm Virtual}$ is the renormalized one-loop amplitude.

We regulate ultraviolet (UV) and infrared (IR) divergences in $d=4-2\epsilon$ dimensions, using $\epsilon_{\mathrm{UV}}$ and $\epsilon_{\mathrm{IR}}$ to identify the origins of the poles. The matrix $\gamma_5$ is treated with the reading-point prescription of Ref.~\cite{Korner:1991sx}.

The tensor one-loop integrals are reduced to the scalar one-loop functions $A_0$, $B_0$, $C_0$, and $D_0$ by tensor decomposition, partial fractioning, and integration-by-parts (IBP) identities~\cite{IBP1, IBP2}. Analytic expressions for their divergent parts are given in Ref.~\cite{Ellis:2007qk}. The dimensional-regularization scale is identified with the renormalization scale $\mu_R$ in the renormalized expressions.

The UV divergences are removed by renormalizing the strong coupling, charm-quark mass, quark field, and gluon field. The bare and renormalized quantities are related by
\begin{align}
	g_s^0 &= Z_g g_s, \notag\\
	m^0_c &= Z_m m_c, \notag\\
	\Psi_0 &= \sqrt{Z_2} \Psi_r, \notag\\
	A_0^\mu &= \sqrt{Z_3} A_r^\mu,
\end{align}
where $Z_i=1+\delta Z_i$ for $i=g,m,2,3$. We use the on-shell (OS) scheme for the charm-quark mass and the quark and gluon fields, and the modified minimal-subtraction ($\overline{\rm MS}$) scheme for $g_s$. The one-loop counterterms are
\begin{align}
		\delta Z_2^{\text{OS}} &= -C_F \frac{\alpha_s}{4\pi} \left[ \frac{1}{\epsilon_{\mathrm{UV}}} + \frac{2}{\epsilon_{\mathrm{IR}}} - 3\gamma_E + 3\ln \frac{4\pi\mu_R^2}{m_c^2} + 4 \right], \notag\\
		\delta Z_m^{\text{OS}} &= -3C_F \frac{\alpha_s}{4\pi} \left[ \frac{1}{\epsilon_{\mathrm{UV}}} - \gamma_E + \ln \frac{4\pi\mu_R^2}{m_c^2} + \frac{4}{3} \right], \notag\\
		\delta Z_3^{\text{OS}} &= \frac{\alpha_s}{4\pi} \left[ (\beta_0' - 2C_A) \left( \frac{1}{\epsilon_{\mathrm{UV}}} - \frac{1}{\epsilon_{\mathrm{IR}}} \right) \right. \notag\\
		&\quad \left. - \frac{4}{3} T_F \left( \frac{1}{\epsilon_{\mathrm{UV}}} - \gamma_E + \ln \frac{4\pi\mu_R^2}{m_c^2} \right) \right], \notag\\
		\delta Z_g^{\overline{\text{MS}}} &= -\frac{\beta_0}{2} \frac{\alpha_s}{4\pi} \left[ \frac{1}{\epsilon_{\mathrm{UV}}} - \gamma_E + \ln(4\pi) \right],
\end{align}
where $\gamma_E$ is the Euler--Mascheroni constant, $\beta_0=11C_A/3-4T_F n_f/3$, and $\beta_0'=\beta_0+4T_F/3$. We use $n_f=4$, with the charm-quark mass retained in the massive loop contribution to $\delta Z_3^{\rm OS}$. For $SU(3)_c$, the color factors are $C_A=3$, $T_F=1/2$, and $C_F=4/3$.

In calculating the virtual corrections, we use the method of regions~\cite{beneke} to extract the NRQCD SDCs directly from the hard-region contributions. The low-energy contributions associated with the projected heavy-quark pair are reproduced by the NRQCD effective theory and drop out of the matching; in particular, the potential-region contribution responsible for Coulomb singularities does not contribute to the SDCs~\cite{Jia:2015quasi}. To isolate the hard contribution, we expand the loop integrands in the relative momentum $q$ before integration. At leading order in $v$, this amounts to setting $q=0$ in the integrands~\cite{Zheng:2022eta}.

\subsection{Real corrections}

\begin{figure}[htb]
    \centering
	\includegraphics[width=0.45\textwidth]{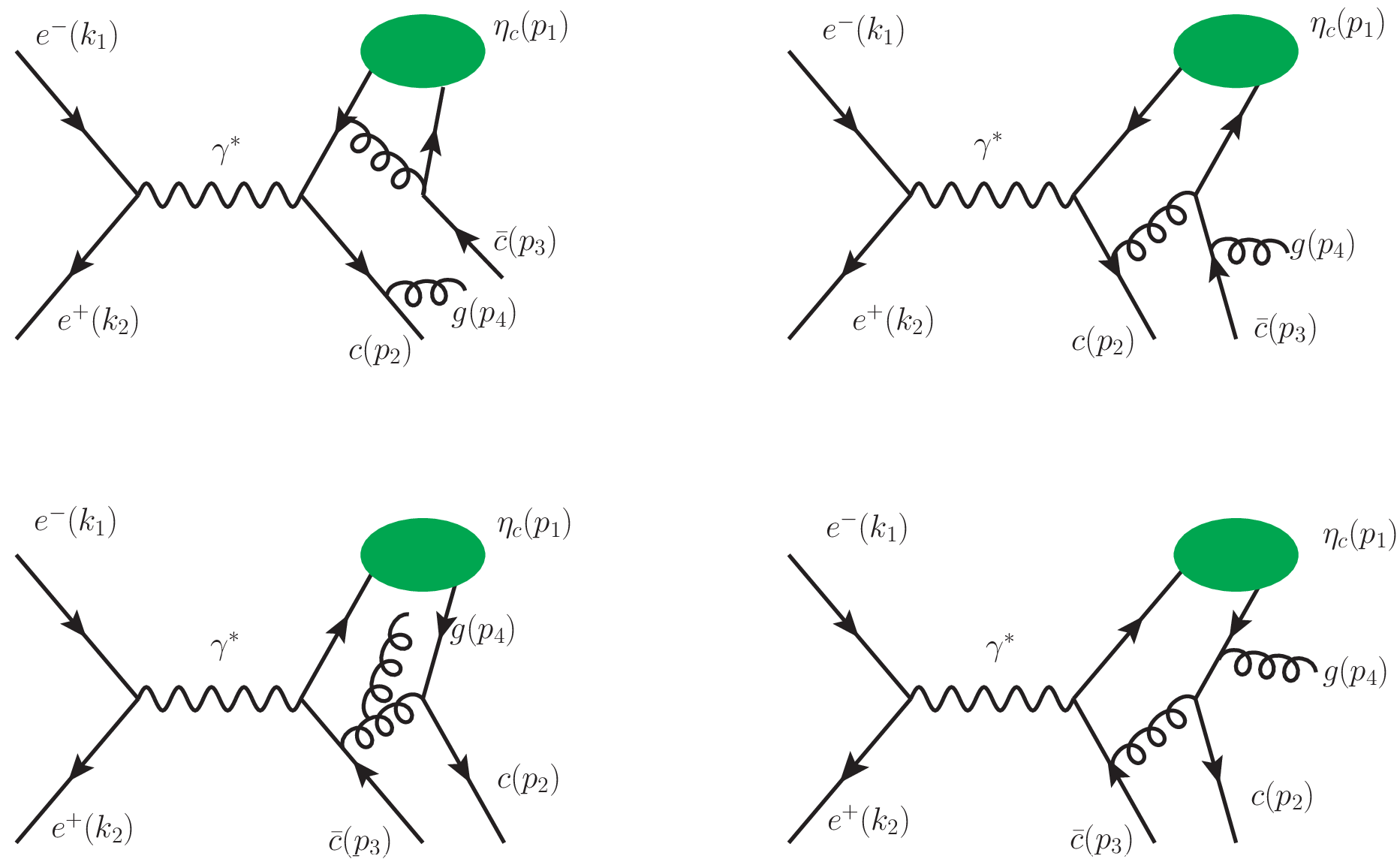}
	\caption{Representative real-emission diagrams for $e^+e^-\to(c\bar c)[{}^1S_0^{[1]}]+c\bar c+g$.
	} \label{real}
\end{figure}

The real contribution is obtained from $e^-(k_1)+e^+(k_2)\to(c\bar c)[n](p_1)+c(p_2)+\bar c(p_3)+g(p_4)$, integrated over the gluon phase space. There are 34 diagrams, with representative examples shown in Fig.~\ref{real}. The corresponding SDC is
\begin{equation}
d\hat{\sigma}_{\rm Real}=\frac{1}{2N_c}\frac{1}{4}\frac{1}{2s}\sum \vert \mathcal{M}_{\rm Real} \vert^2 d\Phi_4,
\end{equation}
where $d\Phi_4$ denotes the differential phase space for the real corrections. 

The real-emission contribution contains soft singularities. Collinear singularities are regulated by the nonzero charm-quark mass $m_c$. The real contribution is evaluated in the same dimensional-regularization convention as the virtual contribution, so that their infrared poles cancel in the inclusive sum.

We isolate the soft singularities with phase-space slicing~\cite{Harris:2001sx}. A dimensionless cutoff $\delta_s\ll1$ divides the gluon phase space into a soft region, $p_4^0\leq\delta_s\sqrt{s}/2$, and a hard region, $p_4^0>\delta_s\sqrt{s}/2$, with $p_4^0$ defined in the $e^+e^-$ center-of-mass frame. Thus,
\begin{equation}
	d\hat{\sigma}_{\mathrm{Real}}
	=d\hat{\sigma}_{\mathrm{Real}}^{\mathrm{soft}}+d\hat{\sigma}_{\mathrm{Real}}^{\mathrm{hard}},
\end{equation}
where
\begin{align}
d\hat{\sigma}_{\mathrm{Real}}^{\mathrm{soft}} \simeq{}& \frac{1}{2N_c}\frac{1}{4}\frac{1}{2s}\notag\\
&\times\left.\sum |\mathcal{M}_{\rm Real}^{\rm soft}|^2d\Phi_4
\right|_{p_4^0\leq\delta_s\sqrt{s}/2},
\end{align}
and
\begin{align}
d\hat{\sigma}_{\mathrm{Real}}^{\mathrm{hard}} ={}& \frac{1}{2N_c}\frac{1}{4}\frac{1}{2s}\notag\\
&\times\left.\sum |\mathcal{M}_{\rm Real}|^2d\Phi_4
\right|_{p_4^0>\delta_s\sqrt{s}/2}.
\end{align}
Here $\mathcal{M}_{\rm Real}^{\rm soft}$ is the leading eikonal approximation to the full real-emission amplitude $\mathcal{M}_{\rm Real}$. The hard contribution uses the full amplitude at the retained order in $v$, with only the phase-space restriction indicated above. It is finite and is integrated numerically in four dimensions. The soft contribution is evaluated analytically in $d$ dimensions, neglecting power corrections that vanish as $\delta_s\to0$.

At $q=0$, soft emission from the two constituents of the color-singlet pair cancels. The remaining eikonal factor involves the open charm quark and antiquark. Neglecting terms suppressed by the soft cutoff, the squared amplitude becomes
\begin{align}
		\sum |\mathcal{M}&_{\text{Real}}^{\text{soft}}|^2 = {}  4\pi\alpha_s C_F \mu^{2\epsilon} \sum |\mathcal{M}_{\text{LO}}|^2 \notag\\
		& \times \Biggl[ -\frac{p_2^2}{(p_2 \cdot p_4)^2} + \frac{2p_2 \cdot p_3}{(p_2 \cdot p_4)(p_3 \cdot p_4)}\notag\\
		&\qquad - \frac{p_3^2}{(p_3 \cdot p_4)^2} \Biggr].
\end{align}
To the same accuracy, the phase space factorizes as
\begin{equation}
\left. d\Phi_4 \right|_{p_4^0 \leq \delta_s \sqrt{s}/2} \simeq d\Phi_3 \left. \frac{d^{d-1}p_4}{2p_4^0 (2\pi)^{d-1}} \right|_{p_4^0 \leq \delta_s \sqrt{s}/2},
\end{equation}
where $d\Phi_3$ is the Born phase space. Integrating the eikonal factor over $p_4$ using Ref.~\cite{Denner:1993a} gives
\begin{equation}
	d\hat{\sigma}_{\text{Real}}^{\text{soft}} \simeq d\hat{\sigma}_{\text{LO}} \frac{\alpha_s}{\pi} \Gamma(1+\epsilon) \left( \frac{4\pi\mu^2}{s} \right)^\epsilon \left( \frac{A}{\epsilon} + B \right), 
\end{equation}
where
\begin{align*}
A={}&C_F\left[1-\frac{\kappa(p_2\cdot p_3)}{(\kappa^2-1)m_c^2}\ln\kappa^2\right],\\
B={}&-A\ln\delta_s^2+C_F\Bigg\{
\frac{1}{2\beta_c}\ln\frac{1+\beta_c}{1-\beta_c}\\
&+\frac{1}{2\beta_{\bar c}}\ln\frac{1+\beta_{\bar c}}{1-\beta_{\bar c}}\\
&+\frac{2\kappa(p_2\cdot p_3)}{(\kappa^2-1)m_c^2}
\big[\mathcal F(\kappa p_2)-\mathcal F(p_3)\big]\Bigg\},
\end{align*}
and
\begin{align*}
\mathcal F(u)={}&\frac14\ln^2\frac{u^0-|\mathbf u|}{u^0+|\mathbf u|}\\
&+\operatorname{Li}_2\!\left(1-\frac{u^0+|\mathbf u|}{\nu}\right)\\
&+\operatorname{Li}_2\!\left(1-\frac{u^0-|\mathbf u|}{\nu}\right).
\end{align*}
Here $\operatorname{Li}_2$ is the dilogarithm, and
\begin{align*}
\beta_c&=\sqrt{1-m_c^2/(p_2^0)^2},\\
\beta_{\bar c}&=\sqrt{1-m_c^2/(p_3^0)^2},\\
\nu&=\frac{(\kappa^2-1)m_c^2}{2(\kappa p_2^0-p_3^0)},\\
\kappa&=\frac{p_2\cdot p_3+\sqrt{(p_2\cdot p_3)^2-m_c^4}}{m_c^2}.
\end{align*}

The infrared pole in the renormalized virtual contribution has the opposite coefficient,
\begin{equation*}
\left.d\hat\sigma_{\rm Virtual}\right|_{1/\epsilon_{\mathrm{IR}}}
=-d\hat\sigma_{\rm LO}\frac{\alpha_s}{\pi}\frac{A}{\epsilon_{\mathrm{IR}}},
\end{equation*}
and cancels the pole in the soft contribution. The cutoff $\delta_s$ is an auxiliary parameter: its dependence must cancel between the soft and hard contributions, up to the power corrections neglected in the slicing approximation.
The stability of the total NLO cross section has been checked by varying $\delta_s$ from $10^{-4}$ to $10^{-6}$.

The calculation outlined above is implemented using the following symbolic and numerical tools. The Feynman diagrams and amplitudes are generated with \textsc{FeynArts}~\cite{Hahn:2000kx}, and the Dirac and color algebra is performed with \textsc{FeynCalc}~\cite{Mertig:1990an, Shtabovenko:2016sxi}. We use \texttt{\$Apart}~\cite{Feng:2012iq} for partial fractioning and \textsc{FIRE}~\cite{Smirnov:2008iw} for the IBP reduction. The scalar one-loop integrals are evaluated with \textsc{LoopTools}~\cite{Hahn:1998yk}, and the phase-space integrations are carried out with \textsc{Vegas}~\cite{Lepage:1977sw}.

\section{Numerical results and discussion}
\label{secNumer}

The direct $\eta_c$ cross sections reported below are obtained by multiplying the SDCs by $\langle {\cal O}^{\eta_c}({}^1S_0^{[1]})\rangle$ as described in Sec.~\ref{secLO}. The input parameters are
\begin{align}
	& m_c=1.5 \pm 0.1\,\text{GeV},\; \sqrt{s}=10.6\,\text{GeV},\;\alpha=1/137,  \notag\\
	& \alpha_s(2m_c)=0.26,\;| R_{\eta_c}(0)|^2=0.81\,\text{GeV}^{3}.
\end{align}
The wavefunction input is taken from Ref.~\cite{originvalue}. The value $\alpha_s(2m_c)=0.26$ corresponds to the central charm-quark mass $m_c=1.5~\mathrm{GeV}$ and agrees with the input in Ref.~\cite{Liu:2004}. The coupling is evaluated with two-loop running using \texttt{RunDec3}~\cite{Herren:2017osy}, with the input $\alpha_s(3.0~\mathrm{GeV})=0.26$. Unless stated otherwise, we use $\mu_R=2m_c$ and quote full-phase-space cross sections. Throughout the tables, $\sigma_{\rm LO}$ denotes the LO term, $\Delta\sigma_{\rm NLO}$ the additive NLO correction, and $\sigma_{\rm NLO}=\sigma_{\rm LO}+\Delta\sigma_{\rm NLO}$ the total through NLO; we define $K\equiv\sigma_{\rm NLO}/\sigma_{\rm LO}$. Here, $\Delta\sigma_{\rm NLO}=\sigma_{\rm Virtual}+\sigma_{\rm Real}$. In the mass scan, $\mu_R=2m_c$ varies with $m_c$, the coupling $\alpha_s(2m_c)$ is evolved accordingly, and $|R_{\eta_c}(0)|^2$ is held fixed.

\subsection{Direct production}
\label{secDirect}

\begin{table}[htb]
\centering
\begin{tabular}{c c c c}
\hline
 $m_c~(\mathrm{GeV})$ \; & $\sigma_{\rm LO}$ \; & $\sigma_{\rm NLO}$ \; & $K$ \\
\hline
1.4 \; & 93.1  & 182 & 2.0 \\
1.5 \; & 58.7  & 118 & 2.0 \\
1.6 \; & 37.1  & 76.8 & 2.1 \\
\hline
\end{tabular}
\caption{Direct production cross sections at $\sqrt{s}=10.6~\mathrm{GeV}$ for the charm-quark masses shown and $\mu_R=2m_c$. Cross sections are in fb; $\sigma_{\rm LO}$ and $\sigma_{\rm NLO}$ denote the LO term and NLO total, respectively.}
\label{tb.etacsection.mc}
\end{table}

\begin{table}[htb]
\centering
\begin{tabular}{c c c c c }
\hline
   Scale & $\sigma_{\rm LO}$ \; & $\Delta\sigma_{\rm NLO}$ \; & $\sigma_{\rm NLO}$ \; & $K$ \\
\hline
$\mu_R = m_c$ &  111 & 81.7 & 193 & 1.7  \\
$\mu_R = 2 m_c$ & 58.7 & 59.3 & 118 & 2.0\\
$\mu_R = 4 m_c$ &  36.5 & 42.9 & 79.3 & 2.2 \\
\hline
\end{tabular}
\caption{Renormalization-scale dependence of the direct production cross section at $\sqrt{s}=10.6~\mathrm{GeV}$ and $m_c=1.5~\mathrm{GeV}$. Cross sections are in fb, with each entry rounded independently; $\Delta\sigma_{\rm NLO}$ is the additive correction and $\sigma_{\rm NLO}=\sigma_{\rm LO}+\Delta\sigma_{\rm NLO}$.}
\label{tb.etacsection.mur}
\end{table}

Tables~\ref{tb.etacsection.mc} and \ref{tb.etacsection.mur} show the charm-mass and the renormalization-scale dependence of direct production. At the central inputs, the LO result is $58.7~\mathrm{fb}$, in agreement with Ref.~\cite{Liu:2004}. The NLO result is $118~\mathrm{fb}$, giving $K\equiv\sigma_{\rm NLO}/\sigma_{\rm LO}\simeq2.0$. This large correction makes the NLO contribution essential for a quantitative prediction.

As $m_c$ increases from $1.4$ to $1.6~\mathrm{GeV}$, both the LO and NLO cross sections decrease substantially; the NLO result falls from $182$ to $76.8~\mathrm{fb}$. A larger charm-quark mass raises the production threshold and reduces the available phase space at fixed collision energy. With $\mu_R=2m_c$, it also increases the renormalization scale, so that the running coupling becomes smaller and further suppresses the rate. The short-distance coefficients also depend explicitly on $m_c$. Despite this substantial variation in the cross section, $K$ changes only from $2.0$ to $2.1$, showing that the relative NLO enhancement is rather stable over the mass range considered.

At fixed $m_c=1.5~\mathrm{GeV}$, we vary $\mu_R$ from $m_c$ to $4m_c$ around the central value $2m_c$. The tabulated extrema give the LO and NLO variations
\begin{align}
36.5~\mathrm{fb}&\leq\sigma_{\rm LO}\leq111~\mathrm{fb},\notag\\
79.3~\mathrm{fb}&\leq\sigma_{\rm NLO}\leq193~\mathrm{fb}.
\label{eq.direct.scale}
\end{align}
Relative to the central prediction at $\mu_R=2m_c$, the LO cross section changes by approximately $+89\%$ and $-38\%$ at $\mu_R=m_c$ and $4m_c$, respectively; at NLO the corresponding changes are $+64\%$ and $-33\%$. The NLO correction remains positive and substantial at all three scales, with $K$ increasing from $1.7$ to $2.2$ as $\mu_R$ increases. As expected, the NLO scale dependence is smaller in relative terms than at LO, but it remains sizable.\footnote{It is noted that as in the $J/\psi$ case~\cite{Huang:2024zni}, the principle of maximum conformality (PMC)~\cite{pmc1,pmc2,pmc3,pmc4,pmc5} also yields, at NLO, a direct-production prediction of $\eta_c$ independent of the choice of renormalization scale, with the PMC scale $Q_*$ determined at leading-logarithmic accuracy. For the central inputs, we obtain $Q_* \simeq 1.5~\mathrm{GeV}$ and $\sigma_{\rm NLO}^{\rm PMC}\simeq193~\mathrm{fb}$, in numerical agreement with the conventional prediction at $\mu_R=m_c$. A detailed PMC analysis is in preparation.}

\subsection{Prompt production with radiative feed-down}
\label{secFeeddown}

In this subsection, we estimate the prompt $\eta_c+c\bar{c}+X$ production by adding to the direct result the dominant radiative feed-down contributions from directly produced $J/\psi$ and $h_c(1P)$. Contributions from higher charmonium states and multistep cascades are expected to be subleading for the total prompt rate. In these cases, $X$ also includes the photon emitted in the radiative decay. In the narrow-width approximation and without acceptance cuts, the integrated feed-down rate equals the parent-production cross section multiplied by the corresponding decay branching fraction.

For prompt production, we fix $m_c=1.5~\mathrm{GeV}$ and evaluate the renormalization-scale uncertainty by varying $\mu_R$ over $[m_c,4m_c]$ around the central value $2m_c$. We use a common scale for the direct and parent-production contributions, keeping the wavefunctions and branching fractions fixed. Unless stated otherwise, the feed-down cross sections quoted below refer to the central scale.

For the $J/\psi$ feed-down contribution, we independently calculate the total cross section and differential distributions for direct color-singlet $e^+e^-\to J/\psi+c\bar c+X$ production at NLO using our own code. The total cross section agrees with the results of Refs.~\cite{Zhang:2006ay, Gong:2009ng, Huang:2024zni} when the same input parameters are used. Using $|R_{J/\psi}(0)|^2=|R_{\eta_c}(0)|^2=0.81~\mathrm{GeV}^3$, together with the other inputs of the present analysis, we obtain NLO parent cross sections of $379$, $257$, and $180~\mathrm{fb}$ at $\mu_R=m_c$, $2m_c$, and $4m_c$, respectively. Taking the branching ratio $\mathcal B(J/\psi\to\eta_c\gamma)=0.0182$~\cite{ParticleDataGroup:2026} as the branching-fraction input gives
\begin{align}
\sigma_{J/\psi}^{\rm feeddown}
&=\sigma(e^+e^-\to J/\psi+c\bar c+X)\mathcal B(J/\psi\to\eta_c\gamma)\notag\\
&\simeq257~\mathrm{fb}\times0.0182\simeq4.68~\mathrm{fb}.
\end{align}

For the $h_c(1P)$ feed-down contribution, we also calculate the total cross section and differential distributions using our own code, retaining the color-singlet ${}^1P_1^{[1]}$ channel of $e^+e^-\to h_c+c\bar c$ production at LO. Our total cross section agrees with Ref.~\cite{Wang:2012} when the same input parameters are used. For the squared first derivative of the radial wavefunction at the origin, we use $|R^{\prime}_{h_c}(0)|^2=0.075~\mathrm{GeV}^5$~\cite{originvalue}. With these inputs, we obtain color-singlet parent cross sections of $25.7$, $13.5$, and $8.43~\mathrm{fb}$ at $\mu_R=m_c$, $2m_c$, and $4m_c$, respectively.\footnote{Using $\langle{\cal O}^{h_c}({}^1S_0^{[8]})\rangle=0.0098~\mathrm{GeV}^3$, as in Ref.~\cite{Wang:2012}, our LO calculation gives a cross section of $0.139~\mathrm{fb}$ for the color-octet ${}^1S_0^{[8]}+c\bar c$ channel at the central inputs, about $1\%$ of the color-singlet result. We therefore neglect this contribution in the present feed-down estimate.} With the branching ratio $\mathcal B(h_c\to\eta_c\gamma)=0.64$~\cite{ParticleDataGroup:2026}, the retained feed-down result is
\begin{align}
\sigma_{h_c}^{\rm feeddown}
&=\sigma(e^+e^-\to h_c+c\bar c+X)\mathcal B(h_c\to\eta_c\gamma)\notag\\
&\simeq13.5~\mathrm{fb}\times0.64\simeq8.64~\mathrm{fb}.
\end{align}

Combining these terms with the direct prediction and varying the common renormalization scale gives
\begin{align}
\sigma_{\eta_c}^{\rm prompt}
&=\sigma_{\eta_c}^{\rm direct}
+\sigma_{J/\psi}^{\rm feeddown}+\sigma_{h_c}^{\rm feeddown}\notag\\
&\simeq131^{+85}_{-43}~\mathrm{fb}.
\label{eq.prompt}
\end{align}
The central value corresponds to $\mu_R=2m_c$, and the uncertainty is obtained by summing all three contributions at each common scale before taking the extrema. The prompt cross section is approximately $216~\mathrm{fb}$ at $\mu_R=m_c$ and $88~\mathrm{fb}$ at $\mu_R=4m_c$. This estimate combines direct $\eta_c$ and parent $J/\psi$ production at NLO with parent $h_c$ production at LO.

At the central scale, the two retained feed-down contributions together increase the direct NLO rate by about $11\%$ ($13~\mathrm{fb}$). Although the $h_c(1P)$ production cross section is much smaller than that of the $J/\psi$, its larger radiative branching fraction makes it the larger of the two feed-down sources. Direct production remains the dominant contribution to the prompt rate throughout the scale range considered.

\subsection{Differential distributions}
\label{secDistributions}

We consider the angular distribution $d\sigma/d\cos\theta$ and the momentum spectrum $d\sigma/dP_{\eta_c}$ in the $e^+e^-$ center-of-mass frame. Here, $\theta$ is the angle between the incoming electron and the outgoing $\eta_c$, and $P_{\eta_c}=|\vec p_1|$.

\begin{figure}[htb]
	\centering
	\includegraphics[width=0.45\textwidth]{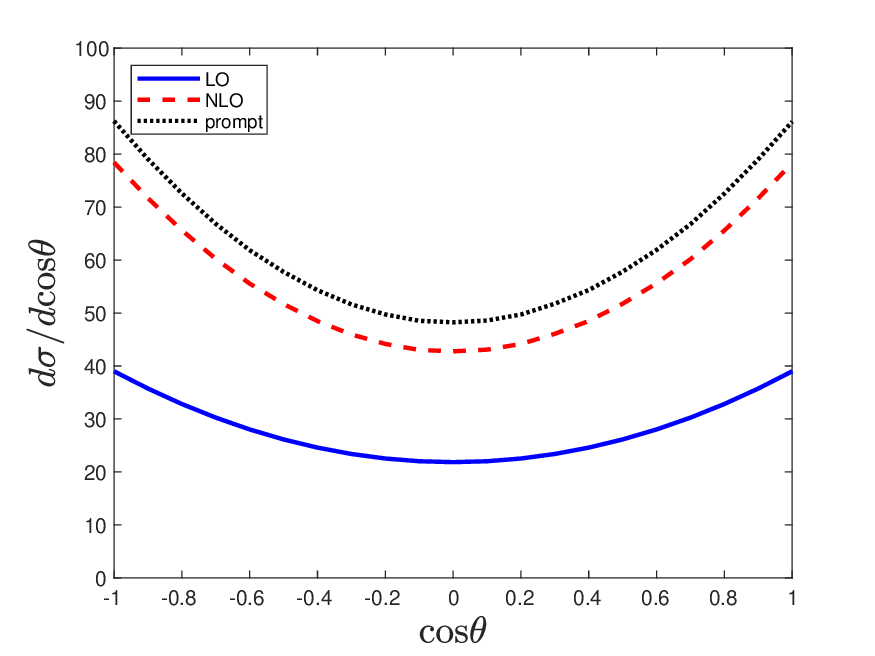}
\caption{Angular distribution $d\sigma/d\cos\theta$ (in fb) at $\sqrt{s}=10.6~\mathrm{GeV}$, $m_c=1.5~\mathrm{GeV}$, and $\mu_R=2m_c$. The solid blue and dashed red curves show direct LO and NLO production; the dotted black curve is the prompt estimate defined in Sec.~\ref{secFeeddown}.}
	\label{fig:cos.distri}
\end{figure}

\begin{figure}[htb]
	\centering
	\includegraphics[width=0.45\textwidth]{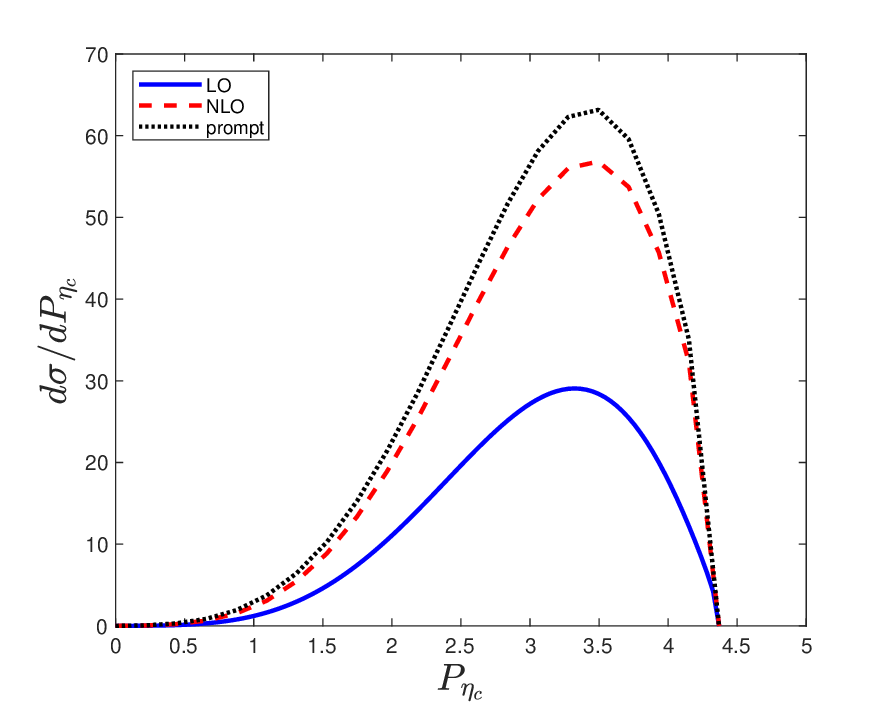}
\caption{Momentum distribution $d\sigma/dP_{\eta_c}$ (in $\mathrm{fb}/\mathrm{GeV}$), with $P_{\eta_c}$ in GeV. The inputs, curve conventions, and treatment of feed-down are the same as in Fig.~\ref{fig:cos.distri}. The NLO curve includes the LO contribution.}
	\label{fig:p1.distri}
\end{figure}

Figures~\ref{fig:cos.distri} and \ref{fig:p1.distri} compare direct LO and NLO production with the prompt estimate of Sec.~\ref{secFeeddown}. We use $m_c=1.5~\mathrm{GeV}$ and $\mu_R=2m_c$. The solid blue, dashed red, and dotted black curves denote direct LO, direct NLO, and estimated prompt production, respectively. Their comparison illustrates the relative importance of the NLO corrections to direct production and the additional radiative feed-down contributions. 

For the feed-down distributions, we neglect decay recoil and approximate the momentum and direction of the daughter $\eta_c$ by those of its parent $H=J/\psi,h_c(1P)$ in the center-of-mass frame: $P_{\eta_c}\simeq P_H$ and $\cos\theta\simeq\cos\theta_H$. The parent distributions calculated in Sec.~\ref{secFeeddown} are then multiplied by the corresponding radiative branching fractions. Writing $x=P_{\eta_c}$ or $\cos\theta$ and $x_H$ for the corresponding parent variable, this gives
\begin{equation*}
\begin{aligned}
\frac{d\sigma_{\eta_c}^{\mathrm{feeddown}}}{dx}
\simeq{}&\sum_{H=J/\psi,h_c}\mathcal B(H\to\eta_c\gamma)\left.\frac{d\sigma_H}{dx_H}\right|_{x_H=x}.
\end{aligned}
\end{equation*}
A similar simplification was used in the $B$-factory study of Ref.~\cite{Gong:2009ng}, where $\psi(2S)$ feed-down was included in the $J/\psi$ momentum and angular distributions through an overall multiplicative factor of $1.29$.

The angular distributions in Fig.~\ref{fig:cos.distri} are symmetric about $\cos\theta=0$, as expected for the unpolarized single-photon channel. All three curves have a U-shaped profile, with larger rates in the forward and backward directions than near $\theta=\pi/2$. NLO corrections enhance the differential cross section by roughly a factor of two over the angular range shown, consistent with the enhancement of the integrated rate. The retained feed-down terms give a further moderate increase while preserving the overall U-shaped profile. The symmetry follows from the production channel, while the U-shaped profile is a result of its dynamics.

The LO, NLO, and prompt momentum spectra in Fig.~\ref{fig:p1.distri} exhibit similar broad, asymmetric profiles with nearby peaks around $P_{\eta_c}=3.3$--$3.5~\mathrm{GeV}$. The spectra rise gradually from small values at low momentum and fall more steeply on the high-momentum side of the peak. The upper kinematic endpoint is set by the minimum invariant mass of the recoiling system. NLO corrections substantially increase the differential cross section across the momentum range shown, and radiative feed-down gives a further moderate enhancement. The enhancement varies with momentum, but the overall peaked shape is largely preserved.

\section{Summary}
\label{secCon}

We have obtained NLO QCD predictions for $e^+e^-\to\gamma^*\to\eta_c+c\bar c+X$ at $B$ factories, including the color-singlet contribution at leading order in the NRQCD velocity expansion. The calculation combines the virtual corrections and real-gluon emission to give an infrared-finite prediction for the direct-production cross section.

At $\sqrt{s}=10.6~\mathrm{GeV}$, $m_c=1.5~\mathrm{GeV}$, and $\mu_R=2m_c$, the direct results and the specified prompt estimate are
\begin{align}
\sigma_{\eta_c,\mathrm{LO}}^{\mathrm{direct}}&=58.7^{+52.3}_{-22.2}~\mathrm{fb},\\
\sigma_{\eta_c,\mathrm{NLO}}^{\mathrm{direct}}&=118^{+75}_{-39}~\mathrm{fb},\\
\sigma_{\eta_c}^{\mathrm{prompt}}&\simeq131^{+85}_{-43}~\mathrm{fb}.
\end{align}
The quoted uncertainties arise from varying $\mu_R$ over $[m_c,4m_c]$ at fixed $m_c=1.5~\mathrm{GeV}$, using a common scale for all prompt contributions. The NLO correction approximately doubles the central direct cross section. The direct prediction also has a substantial mass dependence, spanning $76.8$--$182~\mathrm{fb}$ in the separate mass scan.

The angular distributions retain their symmetric U-shaped profile. The LO, NLO, and prompt momentum spectra have similar overall shapes and nearby peak positions, with substantial enhancement from the NLO corrections and a further increase from feed-down.

The predicted total rate and differential spectra establish an NLO color-singlet baseline for $\eta_c$ production associated with open charm at Belle II. Measurements of this channel can test spin-singlet production with different short-distance dynamics from hadroproduction and complement the constraints from LHCb. These predictions provide the production-level input for comparisons with reconstructed data after the relevant decay and acceptance effects are included. Further refinements include higher-order perturbative and relativistic corrections, together with a more detailed treatment of feed-down.

\begin{acknowledgments}
\noindent\begin{minipage}[b]{\columnwidth}
\setlength{\parindent}{1em}
This work was supported in part by the National Natural Science Foundation of China under Grants No. 12675111, No. 12575080, and No. 12547101.
\end{minipage}
\end{acknowledgments}

\end{document}